\documentclass[aip, apl, 12pt, amsmath,amssymb, reprint]{revtex4-2}

\usepackage{graphicx}
\usepackage{dcolumn}
\usepackage{bm}
\usepackage{amsfonts}
\usepackage{xcolor}

\begin{document}

\title{Vortex Pinning in Niobium covered by a thin polycrystalline Gold Film} 

\author{Wenbin Li}
\affiliation{The Institute for Solid State Physics, The University of Tokyo, 5-1-5 Kashiwa-no-ha, Kashiwa, 277-8581, Japan}

\author{Ivan Villani}
\affiliation{NEST, Istituto Nanoscienze-CNR and Scuola Normale Superiore, Piazza San Silvestro 12, 56127 Pisa, Italy}

\author{Ylea Vlamidis}
\affiliation{NEST, Istituto Nanoscienze-CNR and Scuola Normale Superiore, Piazza San Silvestro 12, 56127 Pisa, Italy}

\author{Matteo Carrega}
\affiliation{CNR-SPIN, Via Dodecaneso 33, 16146 Genova, Italy}

\author{Letizia Ferbel}
\affiliation{NEST, Istituto Nanoscienze-CNR and Scuola Normale Superiore, Piazza San Silvestro 12, 56127 Pisa, Italy}

\author{Leonardo Sabattini}
\affiliation{Center for Nanotechnology Innovation@NEST, Istituto Italiano di Tecnologia, Piazza S. Silvestro 12, 56127 Pisa, Italy}

\author{Antonio Rossi}
\affiliation{Center for Nanotechnology Innovation@NEST, Istituto Italiano di Tecnologia, Piazza S. Silvestro 12, 56127 Pisa, Italy}

\author{Wen Si}
\affiliation{The Institute for Solid State Physics, The University of Tokyo, 5-1-5 Kashiwa-no-ha, Kashiwa, 277-8581, Japan}

\author{Stefano Veronesi}
\affiliation{NEST, Istituto Nanoscienze-CNR and Scuola Normale Superiore, Piazza San Silvestro 12, 56127 Pisa, Italy}

\author{Camilla Coletti}
\affiliation{Center for Nanotechnology Innovation@NEST, Istituto Italiano di Tecnologia, Piazza S. Silvestro 12, 56127 Pisa, Italy}

\author{Sergio Pezzini}
\affiliation{NEST, Istituto Nanoscienze-CNR and Scuola Normale Superiore, Piazza San Silvestro 12, 56127 Pisa, Italy}

\author{Masahiro Haze}
\affiliation{The Institute for Solid State Physics, The University of Tokyo, 5-1-5 Kashiwa-no-ha, Kashiwa, 277-8581, Japan}

\author{Yukio Hasegawa}
\email[]{hasegawa@issp.u-tokyo.ac.jp}
\affiliation{The Institute for Solid State Physics, The University of Tokyo, 5-1-5 Kashiwa-no-ha, Kashiwa, 277-8581, Japan}

\author{Stefan Heun}
\email[]{stefan.heun@nano.cnr.it}
\affiliation{NEST, Istituto Nanoscienze-CNR and Scuola Normale Superiore, Piazza San Silvestro 12, 56127 Pisa, Italy}

\date{\today}

\begin{abstract}
Owing to its superconducting properties, Niobium (Nb) is an excellent candidate material for superconducting electronics and applications in quantum technology. Here we perform scanning tunneling microscopy and spectroscopy experiments on Nb films covered by a thin gold (Au) film. We investigate the minigap structure of the proximitized region and provide evidence for a highly transparent interface between Nb and Au, beneficial for device applications. Imaging of Abrikosov vortices in presence of a perpendicular magnetic field is reported. The data show vortex pinning by the granular structure of the polycrystalline Au film. Our results show robust and homogeneous superconducting properties of thin Nb film in the presence of a gold capping layer. The Au film not only protects the Nb from surface oxidation but also preserves its excellent superconducting properties.
\end{abstract}



\maketitle 

Superconducting electronics is at the heart of the quantum revolution and has triggered recent progress in fields like quantum computing and quantum metrology. Here, the functionality of superconducting devices in the presence of an external magnetic field is crucial for applications. For instance, the superconducting quantum interference device (SQUID) is a very sensitive magnetometer routinely used to measure extremely weak magnetic fields and is based on superconducting loops containing Josephson junctions.\cite{Fagaly2006,Clarke2006,Granata2016,Chieppa2025} Furthermore, several state-of-the-art solid-state qubits are realized with superconducting circuits, and the working point is often set by precise tuning with an external magnetic field.\cite{Krantz2019, Wendin2017, Calzona2022} Another application regards the resonance frequency of flux-driven parametric amplifiers that can be adjusted in-situ via an external magnetic field.\cite{Kutlu2021, Giachero2024, Uchaikin2024} In all the above-mentioned examples, and more, it is thus of fundamental importance to understand the behavior of superconductors in small magnetic fields.

Niobium (Nb) is a very appealing choice for superconducting electronics, for its large superconducting gap, its large critical temperature, and its large critical field. Nb is a type II superconductor (SC), which above a critical magnetic field $H_{c1}$ gets pierced by the magnetic flux field lines that arrange in vortices. Such vortices have been suggested to lead to decoherence in hybrid superconductor-quantum Hall devices in recent studies.\cite{Zhao2023,Schiller2023} Indeed, thanks to high critical field superconductors, the coexistence of superconductivity and quantum Hall regime has recently been demonstrated both in graphene/SC\cite{Amet2016,Vignaud2023,Barrier2024,Villani2025} and III-V semiconductor/SC devices.\cite{Guiducci2019, Hatefipour2022, Hatefipour2024, Akiho2024} Such hybrid devices represent a promising platform for the emergence of topological states. Hence, the study of possible sources of decoherence is important for further advances.

The formation of  vortices in superconductors was visualized and quantified by low-temperature scanning tunneling microscopy and spectroscopy (STM/STS). \cite{Hess1989,Renner1991,Troyanovski1999,Nishizaki2003,Guillamon2009,Cren2009,Ning2010,Cren2011,Yoshizawa2014,Guillamon2014,Roditchev2015,Fente2016,Timmermans2016,Stolyarov2022,Sato2023,Liu2024} Among all, recent work by Odobesko et al.~has been performed on a bare Nb surface prepared and studied under ultra-high vacuum conditions.\cite{Odobesko2019,Odobesko2020} While this work provided evidence for Caroli-de Gennes-Matricon states \cite{Odobesko2019} and anisotropic vortices,\cite{Odobesko2020} the results cannot be directly applied to superconducting devices which are typically exposed to air. To address this issue, the group of Iavarone studied Nb samples exposed to air. They found that the native Nb oxide is too thick to allow for direct tunneling experiments.\cite{Lechner2020} Only after an Ar-ion sputtering could they obtain a surface that allowed for STM operation.\cite{Berti2023}

In this work, we have adopted a different approach and covered the Nb surface with a thin Au film. This is a common approach to protect superconducting samples from oxidation, and since Au does not oxidize under atmospheric conditions, such samples can readily be studied by STM and STS.\cite{Nishizaki2003,Trivini2023} On the other hand, due to the presence of the Au film, which is not an intrinsic superconductor, the situation is more complex, since the superconducting correlations from the Nb film must extend through the Au film to the surface via the proximity effect.\cite{Stolyarov2018,Trivini2023} We add that the capping of superconducting films with a noble metal has very recently been shown to lead to an improvement in transmon qubit coherence, by suppressing surface oxidation.\cite{Bal2024,Chang2025,Ory2025}

We have performed a STM/STS study on Nb islands which were deposited on epitaxial monolayer graphene and protected by a thin film of Au. The epitaxial monolayer graphene was grown on nominally on-axis n-type 6H-SiC(0001) wafers, which were wet-chemically cleaned and hydrogen etched. The graphene growth was obtained by thermal decomposition of the SiC in a Aixtron Black Magic PECVD reactor at high temperature ($1250-1300$~$^\circ$C) under Ar atmosphere\cite{Emtsev2009} employing an AI-assisted adaptive Monte Carlo approach to optimize growth conditions.\cite{Sabattini2025} The Nb islands were defined by electron beam lithography into an array of 2~$\mu$m $\times$ 2~$\mu$m large islands with a pitch of 4~$\mu$m. Then, 30~nm of Nb and 10~nm of Au were sputtered on the graphene. Generally, it is known that such sputtered films are polycrystalline, with a grain size of $30 - 50$~nm.\cite{Gupta2004,Altanany2024} The individual islands were contacted by the graphene film.\cite{Kumar2018} 

\begin{figure}[t]
 \includegraphics[width=\linewidth]{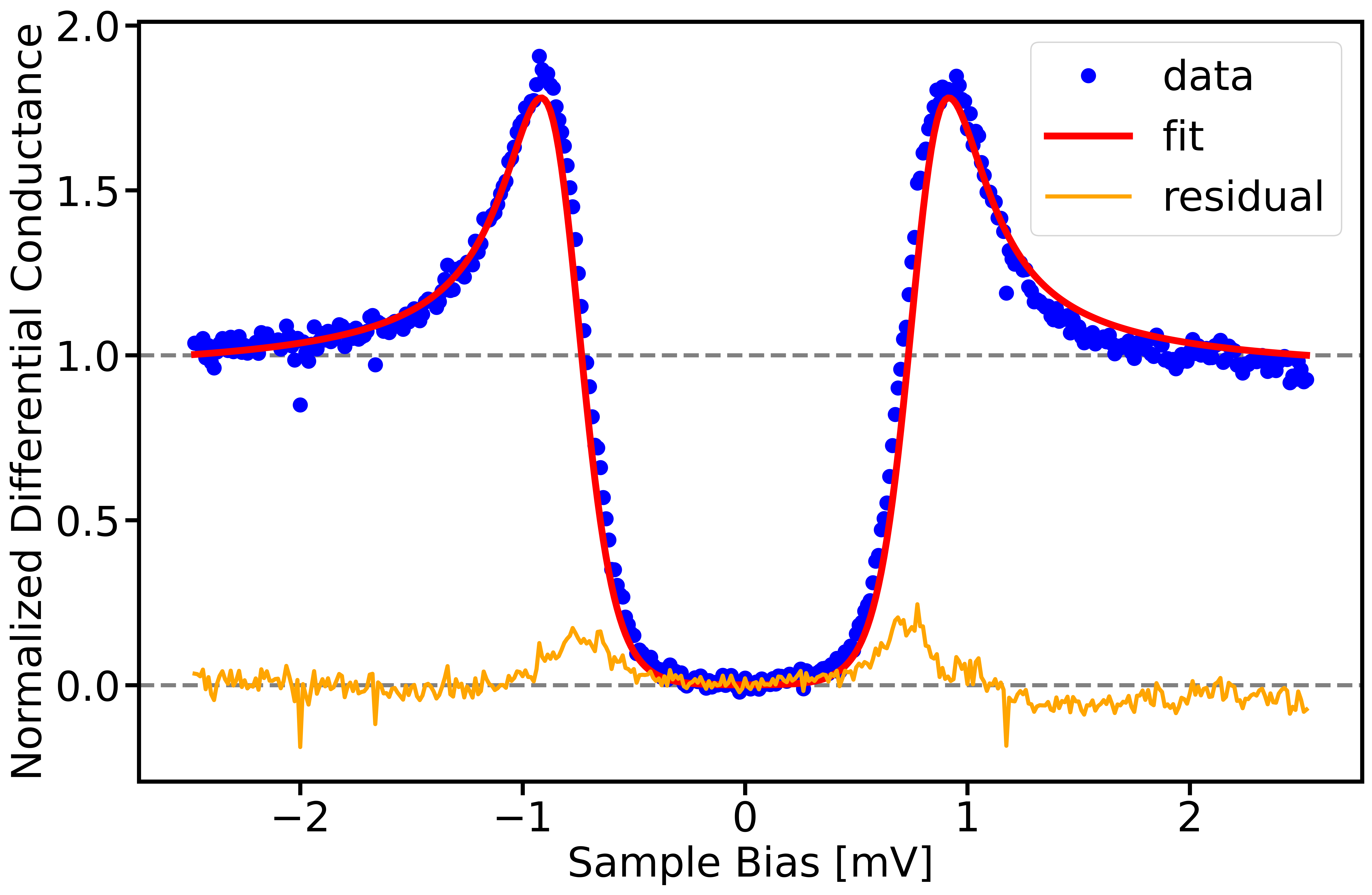}
 \caption{\label{fig1} Normalized differential tunneling conductance of the Nb film for $B = 0$. The data was normalized to the conductance at high bias ($\pm 2.5$~mV). A fit of the data to the BCS-Dynes model is shown, as well (red line), and the corresponding residual (orange line).}
\end{figure}

For Nb films with 60~nm thickness produced under the same experimental conditions (chamber pressure, Nb deposition rate), we previously measured a critical temperature $T_c \sim 9$~K and an upper critical field $H_{c2} \sim 3$~T at $300$~mK.\cite{Guiducci2019} From the value of $H_{c2}$, we can calculate the Ginzburg-Landau coherence length\cite{Eskildsen2002,Ning2010,Sato2023} $\xi = \sqrt{\Phi_0 / (2 \pi H_{c2})}$, with $\Phi_0 = h/(2e)$ the superconducting flux quantum. This gives $\xi = 10.5$~nm, much less than the Nb island size, and in good agreement with literature values for Nb.\cite{Cretinon2005,Pinto2018} Here, instead, we have opted for a 30~nm thick-Nb film, which is the thinnest possible film that already behaves as bulk Nb,\cite{Pinto2018} with a reported $T_c \sim 8$~K.\cite{Pinto2018} This film thickness is reported to show optimized superconducting properties.\cite{Pinto2018} The Nb was protected from oxidation (between deposition and measurements) by a 10-nm-thick Au film, a thickness which guarantees a continuous Au film.\cite{Truscott1999} The elastic mean free path $\ell_e$ of Nb is $\ell_e \sim 5$~nm,\cite{Gupta2004,Pinto2018} which places the samples in the moderately dirty regime ($\ell_e < \xi$).

The STM/STS measurements were performed in a commercial Unisoku USM-1300 system\cite{Nishio2008,Yoshizawa2014,Kim2016a,Sato2023} using (normal metal) PtIr tips. The STM is mounted in a $^3$He cryostat with base temperature $360$~mK. The cryostat is equipped with a 7 T superconducting magnet, whose field is applied perpendicular to the sample. Samples were annealed in ultrahigh vacuum at $160$~$^\circ$C for $10$ minutes and then \textit{in situ} transferred to the STM chamber. The $dI/dV$ spectra, whose intensities correspond to the local density of states (LDOS) of the sample surface, were recorded at a constant STM tip height in the AC lock-in detection mode (modulation frequency 971 Hz, amplitude 50 $\mu$V) by sweeping the sample bias voltage $V_s$. The $dI/dV$ images were taken in the same mode, after the feedback was stabilized at (3~mV, 200~pA) at each pixel point.

Consistent results were obtained from two samples, on a total of 3 islands, and for 9 different applied magnetic field values.

Figure~\ref{fig1} shows a representative normalized differential conductance spectrum obtained on the Au-covered Nb film at zero applied magnetic field ($B = 0$). The spectrum was normalized to the conductance at high bias ($\pm 2.5$~mV). It shows well-developed coherence peaks and a vanishing conductance for small bias. As shown in Fig.~S1 in the Supplementary Material, very similar spectra were measured in different positions of the surface, which demonstrates the homogeneity of the sample. The fact that such spectra could be reproducibly measured shows that the gold film is effective in protecting the Nb film from oxidation.

The data of Fig.~\ref{fig1} was fitted with the BCS model (red line in Fig.~\ref{fig1}) using the Dynes formula for the normalized density of states\cite{Dynes1978}
\begin{equation}
	N_s(E) = N_n \cdot \mathfrak{Re} \left( \frac{E + i \Gamma}{\sqrt{\left( E + i \Gamma \right)^2 - \Delta^2}} \right), \label{eq1}
\end{equation}
with $N_s$ and $N_n$ the density of states in the superconducting and normal state, respectively, $E$ the energy, $\Delta$ the value of the superconducting gap, and $\Gamma$ the phenomenological Dynes parameter which describes the quasiparticle lifetime broadening.\cite{Dynes1978} Thermal broadening is taken into account via the Fermi distribution function $f(E)$\cite{Bergeal2006,Lechner2020}
\begin{equation}
\frac{dI}{dV} \propto -\int^{\infty}_{-\infty} \frac{\partial f(E + eV)}{\partial V} N_s(E) dE. \label{eq2}
\end{equation}
The quality of the fit is high, with a coefficient of determination $R^2 = 0.987$. The fit yields a superconducting gap $\Delta = 0.81$~meV and a very small value for $\Gamma$ compatible with zero. Considering also the finite AC excitation (50~$\mu$V),\cite{Maltezopoulos2003} we obtain an effective electron temperature of the sample of 889 mK. The fact that $\Gamma$ is zero within error bars demonstrates the high quality of the Nb/Au sample and shows that the data can be properly described by the BCS theory, consistent with previous findings.\cite{Moussy2001,Cretinon2005} The value of $\Delta$ is smaller than for bulk Nb, which we attribute to the small Nb film thickness \cite{Moussy2001} and the presence of the Au film.\cite{Truscott1999,Moussy2001} We add that the value of the gap is consistent with previous work.\cite{Moussy2001}

\begin{figure*}[t]
 \includegraphics[width=\linewidth]{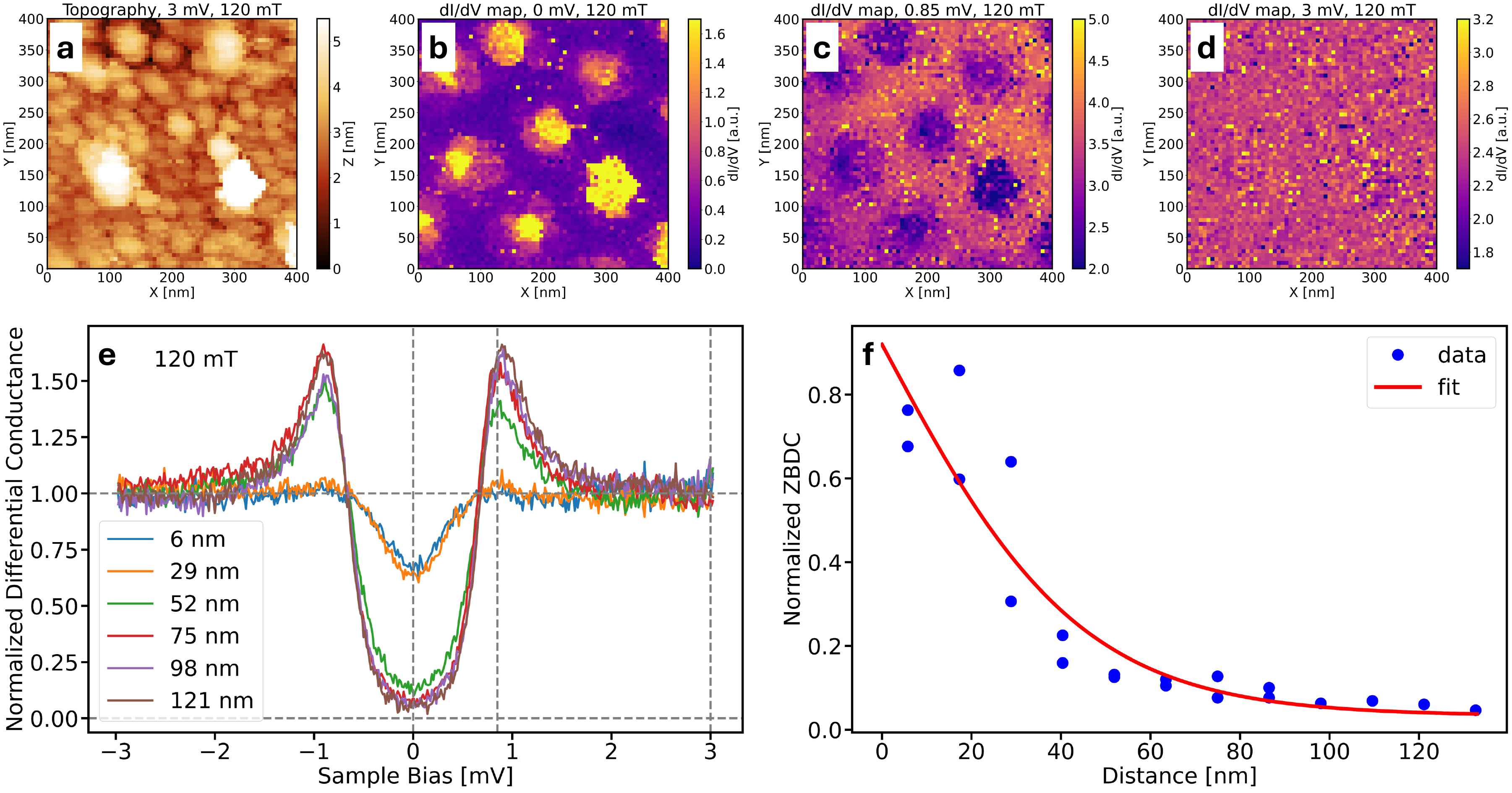}
 \caption{\label{fig2} Vortex imaging at $B = 120$~mT. (a) STM topography of the surface of the Au/Nb film. Scan parameters $U = 3$~mV, $I = 200$~pA. (b-d) Differential conductance maps of the same area as shown in (a) for a DC bias of (b) 0~mV, (c) 0.85~mV, and (d) 3~mV. (a-d) Image size $400$~nm $\times$ $400$~nm. (e) Normalized differential conductance of the film as a function of distance from the vortex core. The vertical dashed lines indicate the bias values at which the data in (b) to (d) was measured. (f) Normalized zero bias differential conductance (ZBDC) as a function of distance from the vortex core. A fit of the data is also shown, from which the superconducting coherence length $\xi$ is obtained.}
\end{figure*}

We have fitted the data also with a more complex theory which was developed especially for proximitized normal materials \cite{Gurevich2017,Kubo2019} and which has been previously applied in the analysis of Nb films covered by a thin oxide layer.\cite{Lechner2020,Berti2023} The results are shown in Fig.~S2 in the Supplementary Material. This additional analysis yielded consistent results, in particular a gap value of $0.78$~meV, and only a slight improvement in $R^2$ to $0.991$. This is a further demonstration of the excellent proximitization of the Au film and thus the high transparency of the  interface between Nb and Au. Therefore, in the following we have analyzed all data with the BCS-Dynes formula (Eq.~(\ref{eq1}) and (\ref{eq2})). 

Next, the effect of a small applied perpendicular magnetic field was investigated. We note that the thickness of the Nb film is much smaller than the London penetration depth ($\lambda_L = 120$ nm for a 30-nm-thick Nb film\cite{Gubin2005}). As shown in Fig.~\ref{fig2}, the differential conductance ($dI/dV$) maps clearly show the presence of Abrikosov vortices at $120$~mT. The position of the vortices approximately follows the expected triangular lattice,\cite{Tinkham2004} but is affected by the morphology of the polycrystalline gold film, shown in Fig.~\ref{fig2}(a) and in Fig.~S3 in the Supplementary Material. At zero bias, as shown in Fig.~\ref{fig2}(b), the vortices appear brighter, i.e., with higher LDOS. On the other hand, at $0.85$~mV, at the position of the coherence peak, they appear darker, as shown in Fig.~\ref{fig2}(c). Finally, for a bias well above the superconducting gap, the map appears homogeneous, see Fig.~\ref{fig2}(d). This shows the homogeneity of the sample and justifies the normalization procedure applied to the spectra of differential conductance. The average distance between vortices is $153 \pm 24$~nm, in good agreement with the theoretical value\cite{Eskildsen2002,Tinkham2004,Bergeal2006,Ning2010} of $d = 1.075 (\Phi_0 / B)^{1/2} = 141$~nm. Their average diameter, evaluated from Fig.~\ref{fig2}(b), is $(36.6 \pm 6.5)$~nm. Repeating the maps at other magnetic fields gives consistent results. Fig.~S4 in the Supplementary Material shows an example for $B = 90$~mT.

\begin{figure*}[t]
 \includegraphics[width=\linewidth]{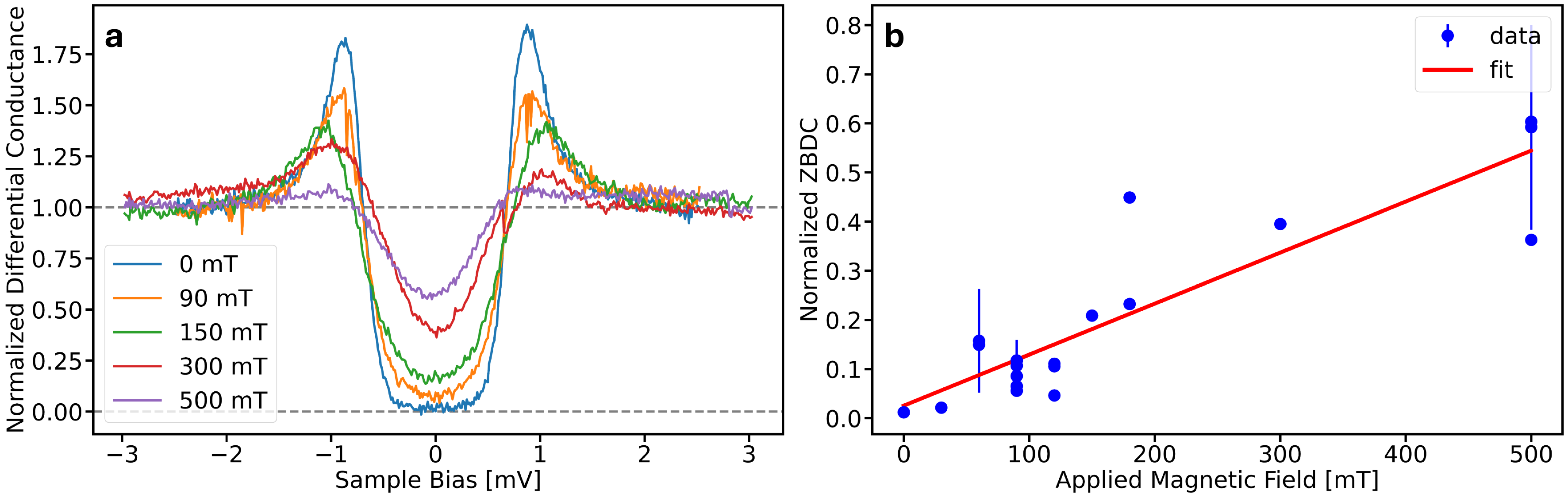}
 \caption{\label{fig3} (a) Variation of normalized differential conductance, measured away from the vortices, with magnetic field, applied perpendicularly to the sample plane. (b) Variation of normalized zero bias differential conductance (ZBDC) with error bars, measured away from the vortices with magnetic fields applied perpendicularly to the sample plane.}
\end{figure*}

The contrast in the maps of differential conductance can be well understood by considering the variation in the normalized differential conductance with the distance from the vortex core. Such data is shown in Fig.~\ref{fig2}(e). At the center of the vortices, the superconducting gap is suppressed but does not vanish, and also the coherence peak is suppressed. This leads to the conclusion that at zero bias, the vortex core has a higher LDOS with respect to regions further away from the vortex core. The opposite is true at the energy of the coherence peak, which explains the inversion in contrast seen in Figs.~\ref{fig2}(b) and (c). Figure~\ref{fig2}(e) also shows that the main variation between on-vortex and off-vortex occurs between 29~nm and 52~nm from the vortex core, in agreement with the reported average diameter of the vortices. On the other hand, no evidence for the presence of Caroli-de Gennes-Matricon states is found in these data, as expected for a Nb film with a mean free path shorter than the coherence length.\cite{Renner1991,Odobesko2020}

The extension of the vortex core into the superconducting film is described by the superconducting coherence length $\xi$. A quantitative measurement of $\xi$ can be obtained by plotting the values of the zero bias differential conductance (ZBDC), i.e., $dI/dV(V=0)$, as a function of distance $r$ from the center of the vortex. This data is shown in Fig.~\ref{fig2}(f). We fit the experimental data with the Ginzburg-Landau expression for the superconducting order parameter,\cite{Eskildsen2002,Bergeal2006,Ning2010,Kim2012} $G = G_0 + A \left( 1 - \tanh \left( -r / \left( \sqrt{2} \, \xi \right) \right) \right)$ (red line in Fig.~\ref{fig2}(f)), with $G_0$ the differential conductance away from the vortex and $A$ a constant. This yields a coherence length $\xi = 31.5$~nm, close to the value of bulk Nb.\cite{Pinto2018} The dimensionless Ginzburg-Landau parameter is $\kappa = \lambda_L / \xi \sim 4 > 1 / \sqrt{2}$,\cite{Tinkham2004,Pinto2018} consistently confirming that this Nb film is a type II superconductor, as expected.

Figure~\ref{fig3}(a) shows the variation in the normalized differential conductance (measured away from the vortices) as a function of applied magnetic field. Increasing the field to $500$~mT decreases the height of the coherence peak and increases the ZBDC to about 0.5, but the induced superconducting gap persists, as expected for these 30-nm-thick Nb films, which have an upper critical field of $H_{c2} \sim 2.3$~T. \cite{Pinto2018} Figure~\ref{fig3}(b) summarizes the data and includes all measured samples. The increase in ZBDC with magnetic field can be fitted with a linear behavior, as expected.\cite{Stolyarov2018}

We expect that the presence of the Au film on top of the Nb film leads to a broadening of the vortices at the surface,\cite{Stolyarov2018} where they are measured by STM. Besides, also the structure of the Au film is expected to affect the nature of the proximity effect.\cite{Truscott1999} Indeed, a visual inspection of Fig.~\ref{fig2}(a) and (b) shows that the vortices are pinned by the grains of the polycrystalline gold film, and preferentially by the larger grains. This is substantiated by the quantitative analysis shown in Fig.~S3 in the Supplementary Material. It has been suggested that the surface height modulation, visible in Fig.~\ref{fig2}(a) and related to the granular structure of the film, leads to a spatial variation of the mean free path in the film, which in turn is responsible for the vortex pinning.\cite{Altanany2024} This observation also provides an explanation why the proximity minigap does not vanish in the vortex cores. It has been reported that the LDOS can sustain a gap when the coherence length $\xi$ becomes comparable to the distance between neighboring grains, i.e., the grain diameter $l$.\cite{Kiselov2023,Melnikov2024} This condition is fulfilled here, with $\xi \approx 0.85 \, l$.

In conclusion, we have investigated the behavior of Nb films which are covered by a thin polycrystalline Au film. The gold film protects the Nb efficiently from surface oxidation. The gold is proximitized, with a sizable minigap, as measured by STS. Applying a perpendicular magnetic field, Abrikosov vortices enter the Nb. Their position is pinned by the grains in the Au film. At the position of the vortex cores, the minigap is reduced but does not completely vanish. We attribute this to the fact that the coherence length is comparable to the grain size. Finally, even at magnetic fields as high as 500 mT, the ZBDC is still about 50\% of the value at zero magnetic field. These features make Nb films protected by thin Au film a very appealing material for applications in quantum technology.

See the supplementary material for data that show the homogeneity of the measured gap, a fit to the theory of Gurevich and Kubo, further evidence for vortex pinning, and additional data at 90 mT.

\begin{acknowledgments}
The stay of S.~Heun at the ISSP of the University of Tokyo was financially supported by the Italian Consiglio Nazionale delle Ricerche in the framework of the Short Term Mobility Program 2023 and 2024. This work is partially supported by the JSPS KAKENHI (Grants Nos.~JP25H00839, JP24K01342, JP22K14598, and JP22H00292). W.~Li acknowledges support by a JSPS postdoc fellowship (No.~P23054). C.~Coletti and A.~Rossi acknowledge financial support from the PNRR MUR Project PE0000023-NQSTI funded by the European Union-NextGenerationEU.
\end{acknowledgments}

\section*{Data Availability Statement}

The data that supports the findings of this study are available within the article [and its supplementary material].

\bibliography{Bib-Nb-vortex}

\end{document}